%% file: insecurity_of_bitcoin.tex
\documentclass[11pt,final,a4paper]{article}

\usepackage[modulo]{lineno}
\usepackage{amsmath, amssymb, amsthm, amsfonts, latexsym, color, paralist,url,ifdraft,lmodern}
\usepackage{thm-restate}
\usepackage[utf8]{inputenc}
\usepackage{csquotes}
\usepackage[T1]{fontenc}
\usepackage[final]{graphicx}
\usepackage{mathtools} 
\mathtoolsset{showonlyrefs}
\usepackage[displaymath,textmath,sections,graphics,floats]{preview}

\usepackage{bbold}
\usepackage{authblk}
\input{macros}
\PreviewMacro*\onote
\PreviewMacro*\nom
\PreviewMacro*\ifdraft

\newcommand{\keygen}{\ensuremath{\mathsf{key\textit{-}gen}}}
\newcommand{\sign}{\ensuremath{\mathsf{sign}}}
\newcommand{\verify}{\ensuremath{\mathsf{verify}}}

\usepackage{booktabs}

\usepackage{algorithm,algorithmicx,algpseudocode}
\usepackage[normalem]{ulem}
\usepackage[draft=false,pageanchor]{hyperref}

\ifdraft{\usepackage{showlabels}}{} %must be after use package{hyperref}.

\newcommand{\ab}[1]{}  %%

\begin{document}
\title{On the insecurity of quantum Bitcoin mining 
\ifdraft{\\(working draft)}
}

\author[1]{Or Sattath}

\affil[1]{Computer Science Department, Ben-Gurion University}

\renewcommand\Authands{ and }
\maketitle
\begin{abstract}
%Short abstract for security and privacy:
%We argue that, in contrast to previous studies, Bitcoin's security is affected by quantum mining, due to an increased stale rate. To circumvent that, we propose a modification to the Bitcoin protocol that discourages the behavior that leads to the increased stale-rate.

Grover's algorithm confers on quantum computers a quadratic advantage over classical computers for searching in an arbitrary data set, a scenario that describes Bitcoin mining. It has previously been argued that the only side-effect of quantum mining would be an increased difficulty. 

In this work, we argue that a crucial argument in the analysis of Bitcoin security breaks down when quantum mining is performed.

Classically, a Bitcoin fork occurs rarely, i.e., when two miners find a block almost simultaneously, due to propagation time effects.  The situation differs dramatically when quantum miners use Grover's algorithm, which repeatedly applies a procedure called a Grover iteration. The chances of finding a block grow quadratically with the number of Grover iterations applied. Crucially, a miner does not have to choose how many iterations to apply in advance. Suppose Alice receives Bob's new block. To maximize her revenue, she should stop and measure her state immediately in the hopes that her block (rather than Bob's) will become part of the longest chain. The strong correlation between the miners' actions and the fact that they all measure their states at the same time may lead to more forks -- which is known to be a security risk for Bitcoin. We propose a mechanism that, we conjecture, will prevent this form of quantum mining, thereby circumventing the high rate of forks.
\end{abstract}

Bitcoin is an electronic payment system~\cite{nakamoto08bitcoin} with a prospective aspiration by its supporters of becoming a dominant form of payment. Likewise, many hope that quantum computers will become an important form of computation. Can these two developments proceed hand-in-hand? Specifically, do quantum computers pose a fundamental threat to Bitcoin's success?

Previous studies that addressed this concern argued that there are no \emph{fundamental} risks to Bitcoin due to quantum computers~\cite{buterin13bitcoin,aggarwal17quantum,tessler17bitcoin}. Their argument is implicitly based on the following false logic: Bitcoin uses two cryptographic primitives, digital signatures and hash functions, both of which have post-quantum variants, i.e., classical digital signatures and hash functions that are secure against quantum adversaries. An upgraded version of this protocol, which will use post-quantum cryptography, will be secure against quantum adversaries. 
In this work, we argue the opposite: a crucial part of Bitcoin's security argument is rendered invalid once Bitcoin miners begin to primarily use quantum computers, even though the cryptographic primitives on which Bitcoin relies on are not broken. This is another instance of what Dominique Unruh calls \emph{the post-quantum fallacy}%
\footnote{See a YouTube video of his talk at \url{https://youtu.be/DgxnNyeWEuE} and the slides are available at \url{https://kodu.ut.ee/~unruh/publications/2016-09-15\%20-\%20Verification\%20of\%20Quantum\%20Cryptography\%20-\%20QCrypt\%20invited\%20talk.pptx}}.

The gist of the argument is as follows. Classically, the times at which miner $A$ and miner $B$ find a block are completely uncorrelated: both continuously try to find a block, and the probability of success for each attempt by either miner is the same and independent of the other miners' actions. This property is tightly related to the fact that proof-of-work is \emph{progress free}~\cite{BK17}. In the quantum setting, this is not the case. Grover's algorithm provides a quadratic advantage to miners, and therefore, the chances of finding a block grow quadratically with the number of Grover iterations (where each iteration takes some fixed time). Suppose Alice devoted 2 minutes to applying Grover's algorithm, and now she receives a new block, mined by Bob. Although she could discard her computation and start mining on top of Bob's block, that course of action would effectively  waste 2 minutes of computational resources. Instead, she could immediately stop Grover's algorithm and measure her quantum state. If she is lucky and her block is valid, and she also propagates her block to most other miners before Bob does, these other miners will mine on top of her block, and she, rather than Bob, will get the block reward. Therefore, it is more profitable for Alice to use the second strategy. We call this second strategy \emph{aggressive}. 

Importantly, as soon as one miner finds a block, all other miners will also measure their quantum states at roughly the same time. In other words, there is a strong correlation between the times at which the different miners measure their respective states and, therefore, the times at which blocks are found. This correlation, which only happens in the quantum setting due to the aforementioned aggressive strategy, leads to a higher fork rate (called the stale rate), which, in the classical setting, is a known security risk. 
\ab{We calculate the stale rate in which all the miners start together, run Grover's algorithm for time $t$ and all measure together \footnote{A realistic setting might be somewhat different: miners would only measure at the same time once they hear about another new block; the starting time (and therefore overall time for the entire protocol) may differ due to previous failures.}
 The stale rate grows approximately linearly in $t$ for values $t<1$ minute (this is not a propagation time effect -- this calculation is done when by taking the limit that the propagation time approaches $0$) and go up to $1$ if $t$ is $10$ minutes. Due to the quadratic speed-up, it seems non-profitable to run Grover's algorithm for times $t\ll 10$ minutes: in the small $t$ regime, the probability that anyone else finds a block is small, and by spending twice the time, yields roughly $4$ times the success probability. Therefore, we conjecture that in equilibrium, the stale rate would be high when miners use this adaptive strategy, unless some countermeasures are taken. 
 }%end of ab

As a countermeasure, we propose to change the default behavior of miners. Currently, in the event of a tie between the longest chains, a miner mines on top of the tip which minimizes the time that the block was received.
We propose to add a term to the tie-breaking rule so as to penalize the aggressive strategy. Our proposed default behavior for the miners is to mine on top of the block, which minimizes the time in which the block was received plus the absolute value of the difference between the time the block was received block and its timestamp.
%\footnote{In case of a tie, different miners may receive the first tip of the competing chains at different times. On the other hand, all miners will agree on the timestamps in the competing blocks they received, since the timestamp is part of the block, and therefore, it cannot be changed.}. 
Greater detail  can be found in the paragraph that immediately precedes Eq.~\eqref{eq:new_tie_breaking_rule}.
The timestamp must be chosen in advance -- before the Grover iterations are applied. From the aggressive miner's perspective, the time at which the competing block will be mined is highly uncertain. Therefore, the timestamps of a block that was mined in response to receiving a competitor's block will not be close to that of the timestamp in its own block. The addition of a term in the default mining behavior will confer a significant penalty for any block created using this aggressive strategy, and therefore prohibit the aggressive strategy usefulness. 

\paragraph{Structure.} First we provide a general explanation, of the cryptographic primitives used in Bitcoin and how the Bitcoin network works. Next the 51\% attack, the stale rate, and the risks associated with a high stale-rate, are discussed. Readers who are familiar with Bitcoin can safely skip this general introduction.
We then provide a basic description of the properties of Grover's algorithm and its relevance to Bitcoin mining. We do not explain \emph{how} Grover's algorithm works, and no prior knowledge in quantum computing is required to follow the text. We then show that the stale rate increases dramatically when miners use a natural quantum strategy. Next, we present our countermeasure to prevent the high stale rate and follow this with our conclusions. Appendix~\ref{sec:subtle-impl-quant} discusses other interesting quantum mining effects that have no implications for the Bitcoin network's security, and Appendix~\ref{sec:unsuccessful_countermeasure} discusses another attempt at devising a countermeasure, but in this case, it was found to have a vulnerability. 

\paragraph{The cryptographic building blocks used in Bitcoin.} 
Bitcoin uses two cryptographic primitives, the first of which is proof of work~\cite{dwork92pricing}. An ideal  cryptographic hash function is modeled as a random function $H:\{0,1\}^*\mapsto \{0,1\}^n$ (this is known as the random oracle model). The function used in Bitcoin (see, e.g.~\cite{narayanan16bitcoin}) is $H(x)=SHA256(SHA256(x))$, for which $n=256$. The challenge in a proof of work is as follows. Given a message $m$ and a target $k$, find an $x$ such that $H(m,x)\leq k$. Because we model $H$ as a random function, the optimal classical algorithm to find $x$ is by brute force.  

The second cryptographic primitive is a digital signature scheme, which is less relevant to our discussion. For a formal definition of digital signatures, see~\cite{goldwasser88digital,katz14introduction,goldreich04foundations}. A signer who wants to sign messages creates a secret and a public key pair, using a polynomial key-generation algorithm, $(sk,pk)\leftarrow \keygen(1^{\lambda})$, where $\lambda$ is the \emph{security parameter}, and the signer publishes $pk$. Using the secret key, the signer can generate a signature $\sigma$ for a document $\alpha$ by running the signing algorithm: $\sigma\leftarrow \sign_{sk}(\alpha)$. A signature thus generated passes verification, that is, \[ \Pr(\verify_{pk}(\alpha,\sigma)=accept)=1.\]
Note that verification can be done by anyone by using the public key of the signer. A forger cannot generate a valid signature in time polynomial in the security parameter $\lambda$. In the chosen message attack, this is formalized by the following security game. The forger has access to a signing oracle $\sign_{sk}(\cdot)$, and the forger wins the security game if he can find a fresh document (i.e., a document that was not given to the signing oracle) and a signature that passes the verification. A signature scheme is unforgeable under a chosen message attack if for every polynomial adversary, the winning probability in this game is negligible, i.e., vanishes faster than $1/poly(\lambda)$.

The current digital signature scheme used in Bitcoin, the Elliptic Curve Digital Signature Algorithm (ECDSA), is not secure against quantum polynomial forgers~\cite{shor97polynomial,roetteler17quantum}. However, digital signature schemes that are designed (and therefore, conjectured) to be secure against quantum forgers -- often called post-quantum digital signature schemes -- already exist. Accordingly, the Bitcoin network should transition to a post-quantum scheme before quantum computers become developed enough to perform such attacks~\cite{buterin13bitcoin,tessler17bitcoin}. Concrete estimates of the time in years before quantum computers will be capable of breaking the elliptical curve signature scheme used by Bitcoin were calculated by Aggarwal et el.~\cite{aggarwal17quantum}. 

\paragraph[A simplistic overview of the Bitcoin protocol]{A simplistic\footnote{We only describe features of the protocol that are relevant to our work. This description does not faithfully represent how Bitcoin actually works. Refer to  Ref.~\cite{narayanan16bitcoin} for a textbook that fully covers the subject. More concise surveys can be found in Refs.~\cite{TS16,Zoh15}.} overview of the Bitcoin protocol.}
Every Bitcoin user creates a digital signature key-pair. Suppose Alice has 5 bitcoins that are already assigned to her secret key $sk_A$, and she wants to send 1 bitcoin to Bob. First, she asks for Bob's Bitcoin address, which is his public key $pk_B$. To send 1 bitcoin to Bob, she creates a message $m=$``I wish to send 1 bitcoin to the address $pk_B$'', and publishes the transaction $tx=(m,\sigma)$, where $\sigma\leftarrow \sign_{sk_A}(m)$. Alice's message, by itself, does not guarantee that Bob will receive the bitcoin: she could create a contradicting message, called a double-spend, where she also sends all $5$ of her bitcoins to Charlie. 

To finalize transactions, the Bitcoin protocol provides incentives to miners -- special nodes in the bitcoin network that use dedicated hardware to solve the proof of work -- to create a valid \emph{block}. If a miner manages to extend the longest (valid) block-chain, she gets a reward of $r$ bitcoins.%
\footnote{The reward $r$, originally set to $50$ bitcoins, is cut in half every 4 years.}
A block consists of transactions, a hash of the previous block, the Bitcoin address of the miner who mined it, the timestamp (which will play an important role later) and a nonce $x$. A block is considered valid if (i) the hash of the block is at most $k$, where $k$ is determined by the blocks that preceded it, and (ii) the transactions are valid, i.e. the signatures pass the verification, and there are no double-spends.

Since each block contains a pointer to a parent, the structure of the block-chain is a (directed) tree, the root of which is called the genesis block. Honest miners try to mine on-top of the tip of the longest block-chain. But what happens if there are two (or more) longest block-chains? The honest (default) behavior is to mine on top of the tip they received first. Currently, these ties occur rarely, and the longest chain rule breaks these ties efficiently, and so, effectively, the graph is a chain (hence the name block-chain), with rare exceptions of short forks.

A miner tries to find a (valid) block by using the proof of work mechanism described above -- the miner increments the nonce $x$ until she finds a valid block that satisfies $H(B,x)<k$. The target $k$ is adjusted by some mechanism (that we do not describe) to obtain  that a block is mined every $10$ minutes in expectation. Currently, the timestamp is part of the block mainly for this adjustment. A user (e.g., a merchant) will consider a received payment as finalized if the transaction in which the bitcoins were transferred is ``buried'' deeply enough inside the longest block-chain. To that end, $6$ blocks built on top of the transaction is considered a high level of security.

\paragraph{The 51\% attack.}
A dishonest miner who controls a fraction of the network can attempt to double spend. Consider the simple attack in which a miner sends one of its coins to a merchant, and after the merchant sends the miner the  goods in return, the miner starts mining on top of the first block that preceded the transaction in which those bitcoins were spent. The attack is considered successful if the longest chain becomes the side of the fork that does not contain the first transaction. The success probability of this attack diminishes exponentially fast as a function of the number of confirmations as long as $q$, the relative hash-rate of the attacker, satisfies $q<\frac{1}{2}$. A 51\% attack refers to the case where $q>\frac{1}{2}$, and here the dishonest miner will eventually succeed with a probability of $1$.  This is one of the main reasons why it is important to maintain Bitcoin mining decentralized.  

\paragraph{Stale blocks.} There are also forks that occur naturally due to network effects, i.e., when two miners mine blocks at almost the same time. 
We call blocks that are outside the longest chain stale% 
\footnote{These blocks are sometimes called orphaned blocks, a term that we feel is confusing: although these blocks do not have descendants, every block, by definition, has a parent indicated by the block to which it is pointing.}
blocks.  We define the stale rate $p_{stale}$ to be the ratio of the number of blocks outside the longest chain to the number of all blocks. The condition $\tau_{prop} \ll \tau_{block}$ is crucial to keep the stale rate small, where the propagation time $\tau_{prop}$ is the time it takes for a block to reach all the other miners after it is mined, and $\tau_{block}$ is the average time between blocks.

Decker and Wattenhofer~\cite{decker13information}  empirically observed a stale rate of 1.69\% in 2013, and, more recently, Stifter et al.~\cite{SSJZKW18} used an improved technique (based on data from merged-mined cryptocurrencies), and estimated that the stale rate between 07/2016 and 07/2018 has dropped to $p_{stale}\approx 0.24\%$.

\paragraph{Concerns with a high stale rate.}
Two main concerns associated with a high stale rate were studied in the classical setting. The first is the increased security risk due to a 51\% attack.
A miner with less than 50\% of the hashing power can double spend by exploiting the 51\% attack described above. Stale blocks are created due to propagation time effects. The attacker can put all of her miners in the same physical location and decrease the propagation time to essentially zero on her own competing chain (it is fairly trivial to arrange a high-bandwidth and low-latency condition, such as in standard data-centers). The condition of this attack being successful (with probability 1) for a miner with a $q$ fraction of the total hashing power is given by:
\begin{equation}
q>(1-p_{stale})(1-q).
\label{eq:double_spend_condition}
\end{equation}
Here, the left hand side is the relative hashing power of the attacker, which is all translated to a chain without any forks. The right hand side is the effective hashing power of the competing (honest) side: though it has a relative hashing power of $1-q$, a $p_{stale}$ fraction of it constitutes stale blocks, and therefore, they do not contribute to the longest chain. Solving Eq.~\eqref{eq:double_spend_condition} gives the condition:
\begin{equation*}
q>\frac{1-p_{stale}}{2-p_{stale}}.
%\label{eq:}
\end{equation*}
For example, if the stale rate is $\frac{1}{3}$, a miner with more than 40\% (instead of 50\%) of the total hashing power can double-spend using this attack.  By plugging in the recent observed stale rate by Stifter et al.~\cite{SSJZKW18} (i.e., 0.24\%) gives the condition $q\gtrsim 0.4994$. This security risk is discussed both theoretically and empirically in more detail in~\cite{decker13information,SSJZKW18}.

The second issue associated with a high stale rate and studied in the classical setting is fairness:  an honest but large miner is systematically rewarded more than her fair share of the total mining revenue (both from the block reward and the fees). Fairness indirectly affects security: unfairness contributes to centralization, and it is considered easier to attack a small group of miners than it is to attack a lot of them (for example, DOS attacks or regulatory changes in few countries) and for the miners to execute a 51\% attack. These concerns are partially addressed in several works~\cite{Sompolinsky15Secure,lewenberg15inclusive,eyal16bitcoin,sompolinsky16spectre,sompolinsky18phantom}. The author is not aware of any PoW-based solution that claims to simultaneously address fairness (of the block rewards and the revenue from fees) and security. 

In the classical vs. the quantum settings, stale blocks emerge for different reasons. In the quantum case, the effects that the stale rate has on security and fairness have not been elucidated. Nonetheless,  we propose a mechanism that we conjecture resolves the high stale rate that may occur as a result of quantum mining. It is not clear, however, whether other concerns also need to be addressed in the quantum setting or whether the counter-measure for the high stale rate also addresses the concerns mentioned above regarding the insecurity and unfairness. 

\paragraph{Grover's algorithm.}

Consider an arbitrary function $f:\{1,\ldots,N\} \mapsto \{0,1\}$. We call an item that is mapped to $1$ by the function $f$ a marked item. If $f$ has $K$ marked items, it takes $\Omega(\frac{N}{K})$ queries to $f$ to find a marked item with probability of at least $\frac{2}{3}$ using a classical random algorithm. Grover's algorithm is a quantum algorithm that finds a marked item with a probability of at least $\frac{2}{3}$ using only $O\left(\sqrt{\frac{N}{K}}\right)$ queries~\cite{grover96fast} (see also the standard text book~\cite{nielsen11quantum}). Furthermore,  $K$ need not be known in advance~\cite{BBHT98}. Grover's algorithm is known to be optimal~\cite{bennett97strengths}.

The above paragraph demonstrates the standard method of presenting Grover's algorithm. Here we offer  a slightly stronger formulation. Suppose we allow only $Q$ queries to the function $f$. Crucially for this work, $Q$ may not be known to the algorithm in advance. The probability of finding a marked item using a classical randomized algorithm is $O(\frac{QK}{N})$, whereas the success probability of Grover's algorithm  is $\Theta(\frac{Q^2 K}{N})$ for $Q=O\left(\sqrt{\frac{N}{K}}\right)$. Grover's algorithm applies a series of identical iterations and can be stopped after any number of iterations. The success probability of finding a marked element when the final state is measured (in the standard basis) is as given above. We emphasize that in the quantum setting, if the outcome is negative, the post-measurement state is of no use (due to the collapse of the quantum state), and one has to start from the beginning to find a marked element. 

\paragraph{Quantum mining.}
A miner that uses a quantum computer can use Grover's quantum algorithm for the function
\begin{equation}
f(x)=\begin{cases}
1, & \mbox{if } H(B,x)\leq k\\
0, & \mbox{otherwise.}
\end{cases}
\label{eq:}
\end{equation}

Consider a miner who receives a new block. In the classical setting, the miner already knows that the work that was invested during the time that elapsed between the previous and the current block was essentially wasted. In the quantum setting, the miner has likely already applied several Grover iterations but has not measured the state. Though the miner could discard that computation, that action would be a waste of resources. Instead, the miner could immediately measure the state and test whether the computation was successful, following a strategy that we refer to as an \emph{aggressive} quantum mining strategy (AQMS -- should be pronounced a-qu-mess as in ``a quantum mess'' ). In contrast, a \emph{peaceful} quantum mining strategy (PQMS) is one in which the state is discarded after receiving a valid block. These two mining strategies are described in more detail in Algorithm~\ref{alg:quantum_mining_strategies}, where the difference can be seen in line~\ref{line:agressive}. We use these terms since an aggressive strategy results in more forks that, in turn, lead to a higher stale rate. 
\begin{algorithm}
\begin{algorithmic}[1]
\algnewcommand\Goto{\textbf{goto}}

\Procedure{Quantum Mining Strategy}{}%{$B_{latest}$}
\State $B\gets$ the tip of the unique longest block-chain, and $k\gets$ the current target
  \Loop
    
    \State \label{line:beginning_of_loop} Create a candidate template for a block $B_{mine}$ without a nonce, with the parent $B$
    \State Define $f(x)=\begin{cases}
1, & \mbox{if } H(B_{mine},x)\leq k\\
0, & \mbox{otherwise.}
\end{cases}$ 
             \State \label{line:sample_Q} Sample $Q$ according to some distribution
	\For{$i=1,\ldots, Q$}
	  \State Apply $1$ Grover iteration, with respect to the function $f$
          \If {a new block $B_{other}$, which is the tip of the unique longest block-chain, is received} 
            \State set $B\gets B_{other}$
            \If{\emph{aggressive}} \label{line:agressive}
              \State \Goto{} line \ref{line:measure}
         \ElsIf{\emph{peaceful}}
              \State \Goto{} line \ref{line:beginning_of_loop}
            \EndIf
          \EndIf
        \EndFor
        \State \label{line:measure} Terminate Grover's algorithm. 
        \If{Grover's algorithm terminated with a successful output $x$   (i.e., an output $x$ for which $f(x)=1$, or alternatively, $H(B_{mine},x)\leq k$)} 
        	\State Set $B\gets (B_{mine},x)$
        	\State \label{line:propagate} Propagate $B$ to all neighbors
        \EndIf
	
  \EndLoop
\EndProcedure
\end{algorithmic}
\caption{Aggressive vs. Peaceful Quantum Mining Strategies}
\label{alg:quantum_mining_strategies}
\end{algorithm}

The analogous aggressive and peaceful \emph{classical} mining strategies are shown in Algorithm~\ref{alg:classical_mining_strategies}. Classically, the difference between aggressive and peaceful strategies in Algorithm~\ref{alg:classical_mining_strategies} only affects the result of one hash per miner per block. Insofar as the current speed of a typical mining device is 10 tera-hashes per second, this distinction between the aggressive and peaceful strategies is too negligible to have a noticeable effect. 

\begin{algorithm}
\begin{algorithmic}[1]
\algnewcommand\Goto{\textbf{goto}}

\Procedure{Classical Mining Strategy}{}%{$B_{latest}$}
\State $B\gets$ the tip of the unique longest block-chain, and $k\gets$ the current target
  \Loop
    
    \State \label{line:classical_beginning_of_loop} Create a candidate template for a block $B_{mine}$ without a nonce, with the parent $B$
    \State Define $f(x)=\begin{cases}
1, & \mbox{if } H(B_{mine},x)\leq k\\
0, & \mbox{otherwise.}
\end{cases}$ 
	\State \sout{Sample $Q$ according to some distribution}
	\State \sout{{\bf for} $i=1,\ldots, Q$}
	  \State \sout{Apply $1$ Grover iteration, with respect to the function $f$} Set $y\gets f(x)$ for a random nonce $x$.
	  \If {a new block $B_{other}$, which is the tip of the unique longest block-chain, is received} 
            \State set $B\gets B_{other}$
            \If{\emph{aggressive}} \label{line:classical_agressive}
              \State \Goto{} line \ref{line:classical_test_success}
         \ElsIf{\emph{peaceful}}
              \State \Goto{} line \ref{line:classical_beginning_of_loop}
            \EndIf
          \EndIf
	\State \sout{\bf{end for}}

        \State \label{line:classical_measure} \sout{Terminate Grover's algorithm.} 
        \State \sout{\bf{if} Grover's algorithm terminated with a successful output $x$   (i.e., an output $x$ for which $f(x)=1$, or alternatively, $H(B_{mine},x)\leq k$)} 
        \If {$y=1$ (i.e. we found a valid block with a nonce $x$)} \label{line:classical_test_success}
        	\State Set $B\gets (B_{mine},x)$
        	\State \label{line:classical_propagate} Propagate $B$ to all neighbors
        \EndIf 
	
  \EndLoop
\EndProcedure
\end{algorithmic}
\caption{Aggressive vs. Peaceful \emph{Classical} Mining Strategies}
\label{alg:classical_mining_strategies}
\end{algorithm}

\paragraph{Analysis of the stale rate in a simplistic model.} We now provide a simplistic model with which we can analyze the stale rate. Suppose there are $n$ symmetric miners (i.e. each miner has exactly the same hardware, software, network facilities, etc.) that are all interconnected with each other (the topology is a clique). Furthermore, we assume that the network is fully synchronous, and we ignore network effects. Note that in a fully synchronous setting with clique topology, there is no point in using an aggressive strategy, since by the time you hear about the other block, all the other nodes have already received it, and no other miner will mine on top of it because they would receive it only \emph{after} the original block (recall the tie-breaking rule). Therefore, we can assume that in both the classical and the quantum settings, all miners use a peaceful strategy.

 In the classical setting, let $h$ be the hashes-per-minute that every miner applies, and $p_{success}=\frac{k}{N}$ be the probability that the value of a hash would be at most $k$ and, therefore, a successful guess. Since a block is mined (roughly) every $10$ minutes in expectation, we deduce that $p_{success}\cdot h \cdot  n\approx \frac{1}{10}$. Suppose one miner finds a block. The probability that another miner will find a block in exactly the same round (and hence, a fork will occur) is at most $n\cdot p_{suceess}\approx \frac{1}{10 h}\approx 1.6\mathrm{e}{-13}$, which can be safely approximated to $0$.

Suppose that in the quantum setting, all miners choose the same $Q$ in Algorithm~\ref{alg:quantum_mining_strategies}, line ~\ref{line:sample_Q} deterministically.  Let us denote by $t$ the number of minutes it takes to apply these $Q$ iterations (i.e., if the time it takes to perform one iteration in the for loop is $t_{iteration}$, then $t=Q\cdot t_{iteration}$). Every $t$ minutes, there is a chance that a block will be mined. We further assume that $n$ is large, and therefore, the number of miners that find a block has a Poisson distribution, and we denote the expectation of it as $\lambda(t)$. Let $E$ be the event that the longest chain gets extended at time $t$, and $F$ be the event that no block is found at time $t$. Note that $F$ is the complement event of $E$, and therefore, the longest chain is extended precisely when there is at least one miner that finds a block: 
\begin{equation}
  \label{eq:1}
  \Pr(E)=1-\Pr(F)=1-e^{-\lambda(t)}
\end{equation}
The Bitcoin network adjusts the difficulty so that the longest chain gets extended every $10$ minutes in expectation. Thus, we can assume:
\begin{align}
  \label{eq:2}
  \frac{1}{10}&=\E[\text{\#blocks added to the longest chain per minute} ]\\
&=\frac{1-e^{-\lambda(t)}}{t}
\end{align}
Therefore, for $0<t<10$,
\begin{equation}
  \label{eq:3}
   \lambda(t)=\ln(\frac{10}{10-t})
\end{equation}
\begin{align}
\label{eq:4}
  \E[\text{\#blocks added per minute} ]=\lambda(t)/t
\end{align}

The stale rate is:
\begin{align}
  \label{eq:5}
  p_{stale}(t)&=\frac{ \E[\text{\#blocks outside the longest chain added per minute} ]}{  \E[\text{\#blocks added per minute} ]}\\
           &=1-\frac{\E[\text{\#blocks added to the longest chain per minute} ]}{ \E[\text{\#blocks added per minute} ]}\\
           &=1-\frac{1/10}{\lambda(t)/t}=1-\frac{t}{10 \ln(\frac{10}{10-t} )}           
\end{align}

\begin{figure}[htp]
\includegraphics[width=0.9\textwidth, height=8cm]{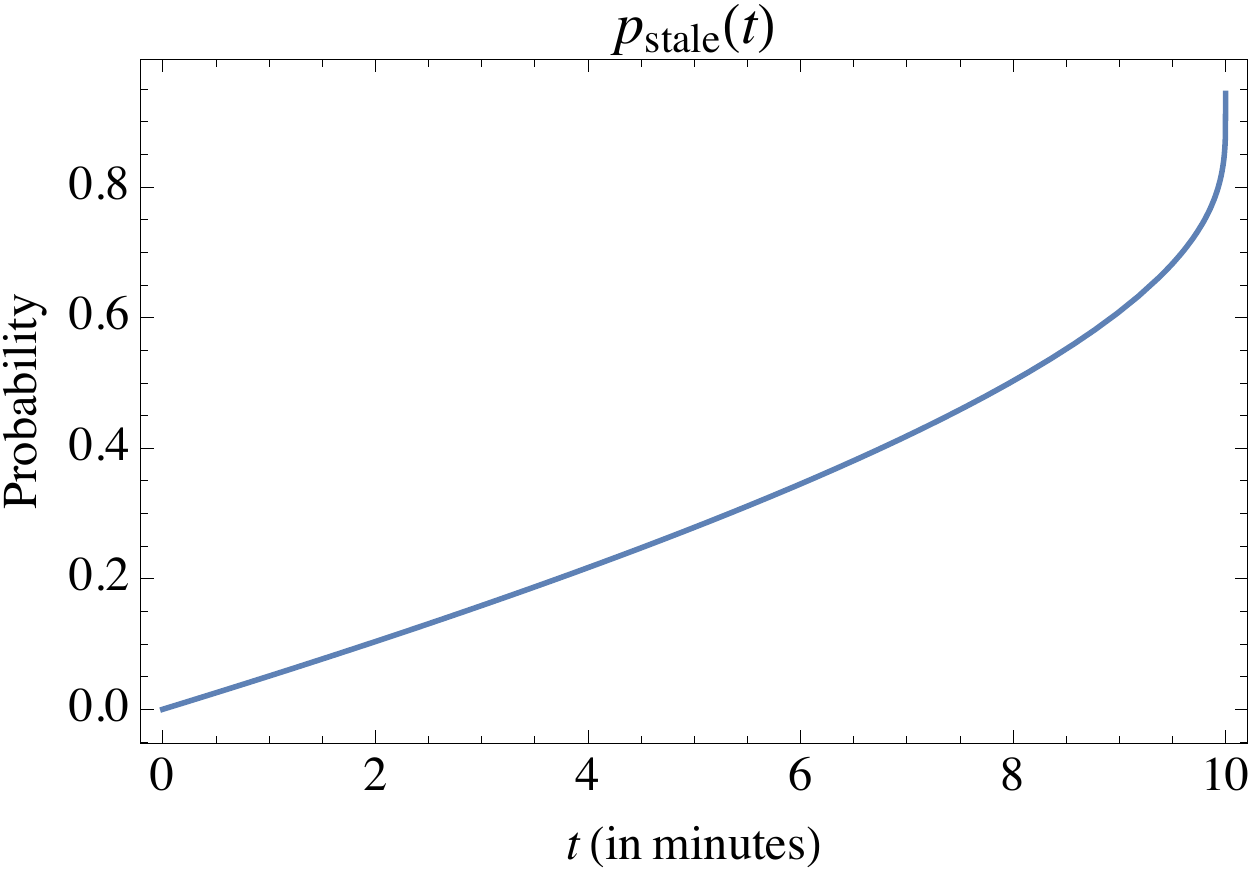}
\caption{The stale rate $p_{stale}$ as a function of the time (measured in minutes) dedicated to perform Grover's algorithm $t$. We assume that, among other things, the number of miners is large and that they all measure at the same time $t$.}
\label{fig:p_stale}
\end{figure}

Figure~\ref{fig:p_stale} presents $p_{stale}(t)$.

For example, the stale rate is $\approx 5\%$ for $t=1$ minute and $50\%$ for $t \approx 7.96$ minutes. 

It may be the case that a deterministic choice of $Q$ is not an equilibrium and that only mixed equilibria strategies (i.e., distributions over $Q$) exist. Note that there are two contradicting forces at work here: by investing more time a miner can increase her chances of  finding a valid nonce quadratically (due to the Grover speed-up). On the other hand, the more time the miner invests in a particular computation increases the risk that someone else will find a block before she does, in which case the time she spent was wasted as she will have to terminate before completion; in other words, investing less time increases the chances that her computation will not be discarded. We restate that in the \emph{fully synchronous} setting, the AQMS has no advantage over the PQMS: by the time a miner hears about another miner's block, all other miners have also received that block, and therefore, there is no chance of winning such a block race.

\paragraph{The stale rate in a realistic setting.} In more realistic scenarios, where propagation time effects are taken into account (i.e., the non-synchronous setting), it is more profitable for quantum miners to use aggressive rather than peaceful strategies.
%\footnote{We  argue that when all other miners use PQMS, then it is better to use an AQMS over a PQMS. It is \emph{not} true that an AQMS dominates the PQMS (that is, regardless of what strategy others are using, the AQMS is no worse than the PQMS). For example, if a powerful miner runs a 51\% attack, it is better to join the attacker's branch (since other branches will eventually loose), and PQMS, implicitly, does that. In fact, to discuss dominant strategies, one has to the define the set of possible strategies first. The class of relevant classical mining strategies is surprisingly rich, and is still a topic of active research (see, e.g.,~\cite{nayak16stubborn_mining}).}
 The second force changes, and depends on the relative fraction of the mining network to which the miner can propagate her block before her competitor, often called the propagation parameter $\gamma$, which plays an important role in other works, for example~\cite{eyal14majority,sapirshtein15optimal,nayak16stubborn_mining}. In case there is a fork, the $\gamma$ parameter of a miner determines the probability that other miners will receive her block first, and according to the current tie breaking rule, she will be the one to receive the block reward by winning the block race. A miner who chooses a lower value for $Q$ decreases the probability of a fork and increases her expected revenue.

%One can model the setting in which all miners use the AQMS by a one shot game in which the set of strategies is compact\footnote{For that, one has to use a continuous variant for Grover search, such as in Ref.~\cite{farhi98analog}, in which the miners choose a continuous parameter $t$ after which they stop and measure. There is always a finite point which guarantees a success probability $1$, so the set of valid times is bounded, and hence compact.} and the utility function is continuous\footnote{Unlike in the context of PQMS, if all users use an AQMS, the utility function is continuous -- there is no significant advantage in increasing or decreasing the time to measure, since in any case, if you are lucky, all other miners who haven't measure will also measure.}. In such games, under mild assumptions, %a \emph{pure} Nash-equilibrium is guaranteed to exist~\cite[Theorem 1.2]{fudenberg91game_theory}
%a pure symmetric Nash equilibrium exists~\cite{CRVW04}. Since such an equilibrium strategy is pure and \emph{symmetric}, all the miners measure at the same time. We cannot rule out a scenario where there is a pure symmetric Nash-equilibrium, in which a large fraction of the miners find a block, and split the reward.  In such a scenario, the Bitcoin network completely loses its ability to break ties in case of a double-spend, which is arguably its main task. 

The main observation is that it is more profitable to use aggressive strategies in a realistic setting (i.e, in the non-synchronous model). All miners will therefore do so, and when one miner finds a block, the rest of the miners immediately measure their states. As a result, there is strong correlation between the times at which valid blocks are found. This outcome is in sharp contrast to the classical behavior, where the distribution of finding a block by two miners is statistically independent. The conclusion that can be drawn from the synchronous model is that if all quantum miners measure at the same time, we should expect a high stale rate, as in Fig.~\ref{fig:p_stale}. 
It is beyond the scope of this work, however, to analyze the properties of equilibria in various models.

It is plausible that one can formulate a one-shot multi-player game that captures many of the behaviors of the properties that we discussed.
\begin{openproblem}[Informal] 
Let $\mathcal{P}$ be the pure PQMSs,  and let $\mathcal{A}$ be the pure AQMSs. We denote by $NE_{P}$ the set of mixed Nash-equilibria when the players can choose any strategy from $\mathcal{P}$, and $NE_{P\cup A}$ is the set of all mixed Nash-Equilibria when the players can choose any strategy from $\mathcal{P} \cup \mathcal{A}$.
 
In all of the open questions, we are interested in the Nash-equilibrium in the setting where the propagation delay is finite but goes to $0$, all miners are symmetric and quantum, and the number of players $n\rightarrow \infty$.
  \begin{enumerate} 
\item \label{it:non-zero-stale-rate} Characterize the Nash-equilibria in $NE_{P \cup A}$.
What are their stale rates? Specifically, are all stale rates $\Omega(1)$? We conjecture that the answer is yes.
 \item \label{it:peaceful_implies_zero_rate} 
Characterize the Nash-equilibria in $NE_{P}$.
What are their stale rates? Specifically, are all stale rates $o(1)$? We conjecture that the answer is yes.
  \end{enumerate}
\label{op:quantum_mining}
\end{openproblem} 
The conjectures in Open Problem~\ref{op:quantum_mining} mean that there are no low stale rate equilibria if AQMSs are allowed and that, when only PQMSs are considered, there exists (where here we use Nash's Theorem~\cite{nash50equilibrium}) a low stale rate equilibrium. Therefore, these conjectures suggest that one way to resolve the high stale rate is to change the Bitcoin protocol such that AQMSs are prohibited. In the next section, we discuss such a countermeasure.  

In a recent paper that was published after the first version of this work, Ray, Lee and Santha~\cite{LRS19} proved a weak variant of our second conjecture. They assume that miners only use the PQMS, an assumption that they justify by exploiting the countermeasure discussed below. They define a game that models some of the central aspects of quantum mining while ignoring others\footnote{
	Specifically, they ignore the fact that once Grover's algorithm fails, the miner restarts; and that after a Bitcoin block is found, the process is repeated. 
}.
They analyze two games, called a quantum race and a stingy quantum race. In the latter case, in case of a fork, all parties loose. In the (non-stingy) quantum race, all the miners that generated the fork split the revenue from the block reward. For technical reasons, the stingy game is easier to analyze, yet, the non-stingy race presents a better model for Bitcoin. They show that when the stingy quantum race is in equilibrium, the probability of a fork goes to $0$ as the hashing power increases -- see~\cite[Theorem 27 and Theorem 8]{LRS19}. They also show that the same strategy is an approximate Nash equilibrium in the (non-stingy) quantum race. Since this is the same strategy, it shows that there exists an \emph{approximate} Nash equilibrium in which the stale rate goes to $0$. This is in line with our conjecture.   

\paragraph{A countermeasure to the aggressive strategy.}
To prohibit use of the AQMS, consider the following proposal. Every Bitcoin block already contains a timestamp. Currently, the timestamp is used mainly to calculate the new target $k$ every 2016 blocks ($\approx 2$ weeks).
We propose to change the default tie-breaking rule. Suppose these latest blocks (each one is the tip of a longest chain) have timestamps $s_1,\ldots, s_n$, and they were first received at the times $t_1,\ldots,t_n$.
%  Let $t_{min}=\min_{i\in[n]} t_i$ and let the score of each of these tips be defined as $\Delta_i=|s_i-t_{min}|$.
For each tip, let $\Delta_i \equiv |s_i-t_i|$ and let the penalty be defined as $p_i\equiv t_i + \Delta_i$.
The current default strategy is to mine ``on-top'' of the tip that was received first, i.e., the block that minimizes $t_i$ (see, for example~\cite[Chapter 5.5]{narayanan16bitcoin}). Instead, we propose that the default strategy is to mine ``on-top'' of the tip that minimizes the penalty $p_i$, i.e., add the term $\Delta_i$ to the penalty calculation. To conclude, the default mining strategy in the event of ties between longest chains should be changed to mining on top of the tip that achieves:
 \begin{equation}
 	\min_i p_i =\min_i t_i + |s_i - t_i|
 	\label{eq:new_tie_breaking_rule}
 \end{equation}

We now explain why this countermeasure is useful. An honest miner knows in advance how many Grover iterations it intends to apply, and under the (very plausible) assumption that she knows how long it will take to complete these iterations, she can set the timestamp to be that (future) time. Therefore, the tip of an honest miner will have a penalty  $p_i$ that is $t_i$ plus $\Delta_i$, where $\Delta_i$ will be the propagation time -- which is an order of a second in practice. 
On the other hand, non-honest miners do not know in advance when their competitors will find a block. Since they have to set the timestamp in advance (and it cannot be altered later, due to the properties of Grover's algorithm and Hash functions; therefore, the timestamp can be viewed as a value to which the miner \emph{commits} at the beginning of Grover's algorithm), most likely this uncertainty would cause the $\Delta_i$ of their blocks to be large.

This change does not require users to upgrade at all -- only miners would be asked to upgrade. Furthermore, this upgrade does not require a hard-fork, and not even a soft-fork (as no damage occurs to non-upgrading miners).  For more detail on these two types of forks, see~\cite[Chapter 3]{narayanan16bitcoin}.  

The strategy that we propose also has some drawbacks. Selfish mining is a strategy that allows a miner (or a mining pool) to obtain more than its fair share~\cite{eyal14majority}. To partially mitigate selfish mining attacks, Eyal and Sirer suggested that for the case of competing longest chains, the default (honest) behavior should be  to pick the tip of the chain to be mined on top of uniformly at random~\cite[Section 6]{eyal14majority}. Another mechanism, called Freshness Preferred, was introduced by Heilman~\cite{Hei14}, whose goal was also to increase the minimal hash-power needed to perform a selfish mining attack. In addition,  the properties of Heilman's mechanism were better than those contained in the proposal of Eyal \& Sirer. Heilman's mechanism uses the timestamp of the block to change the tie-breaking rule, but it does so in a slightly different fashion than we propose. Although it would be convenient if there were a way to mitigate both the problems of selfish mining and the AQMS by using the same mechanism, neither the proposal of Eyal \& Sirer nor that of Heilman is compatible with ours.
The lack of a single approach to these two problems may ultimately force the Bitcoin community to choose which issue is more important: selfish-mining attacks (and in that case, use their proposed solution to mitigate selfish mining attack vectors) or prohibiting aggressive quantum mining (and use our proposed solution). It seems plausible that our proposed mechanism can mitigate both of these problems, since it achieves the design goals set by Heilman (which improved on those of Eyal \& Sirer). However, it is beyond the scope of this work to address the issue of selfish mining and its variants~\cite{sapirshtein15optimal,nayak16stubborn_mining}.

Note that Eq.~\eqref{eq:new_tie_breaking_rule} asks the miner to behave somewhat irrationally: a miner who began to run Grover's algorithm on one tip is asked to move to another tip if the latter has a lower penalty. This means that the miner looses the computational power invested between the time she started and the time she moves to a new tip. Since the AQMS will accrue a high penalty, we expect this to be a rare event, similar to the stale rate today. And even if it does happen, the time spent mining the wrong tip is short (upper-bounded by the propagation time $\Delta_i$), so the damage incurred from moving to the other tip is relatively small. Based on the stale rate in $2017$, this translates into potential damage of roughly one minute every year. This is in contrast to damage of following the PQMS without our proposed countermeasure, which amounts to minutes \emph{per block}, and sum up to several months per year.

Another drawback of our countermeasure is that it opens new attack vectors, such as those based on timing attacks (for example, attacks on time-servers), since the strategy depends on having an accurate clock to calculate $p_1,\ldots,p_n$. As mentioned before, the current Bitcoin protocol barely uses time, and therefore, the Bitcoin network is currently less prone to such attacks. 

An alternative countermeasure approach would be to find a PoW mechanism for which quantum computers have no advantage over classical computers. Aggarwal et al. suggested an alternative proof-of-work function, called Momentum, for which the quantum advantage is claimed to be smaller than the quadratic advantage for double-SHA256 (which is the one used in Bitcoin)~\cite{aggarwal17quantum}. However, their construction is insufficient to completely resolve the security risk we discussed, since there is still a polynomial quantum speed-up. Other approaches try to replace proof of work with an entirely different procedure -- one notable attempt is called proof of stake, which is not prone to the vulnerability discussed in this work. More details on proof of stake can be found in~\cite{KRDO17} and references therein.  

\begin{conjecture} If our proposed countermeasure is deployed, the AQMS becomes ineffective: in the setting as in Open Problem~\ref{op:quantum_mining}, PQMS with $t$ iterations weakly dominates AQMS with $t$ iterations. 
%$NE_{P \cup A}=NE_P$, i.e. all Nash-equilibria in $NE_{P\cup A}$ have support only on the peaceful strategies, $PQMS$. 
\label{con:NE_PA_equal_NE_P}
\end{conjecture}

In conjunction with the second conjecture in Open Problem~\ref{op:quantum_mining}, these conjectures imply that the effective stale rate will be $o(1)$. 
 
 We note that it is not clear whether the security risks that were discussed due to a high stale rate, namely, a reduced threshold for 51\% attacks and unfairness, are relevant in the quantum mining setting. The reasons for stale blocks in the classical and quantum settings (with or without the countermeasure) are different: in the classical setting, a miner can reduce its stale rate on its side of the chain by putting all of its mining equipment in the same physical location. Yet, in the quantum setting, a single miner might also suffer from a high stale rate. Therefore, even if the conjecture above does not hold, nevertheless, it is possible that neither fairness nor security will be affected by quantum mining. 

We leave the issue of  whether security or fairness can be compromised with our proposed counter-measure as an open question. We emphasize that this is a very different question than Conjecture~\ref{con:NE_PA_equal_NE_P} and that it is harder to formalize for various reasons: in reality, miners are not restricted to the AQMS or PQMS, but rather, they can use other strategies; indeed, their goal may be to cause damage to others rather than to increase their own utility (see the example in Appendix~\ref{sec:unsuccessful_countermeasure}); Bitcoin mining is a repeated game, played for each block, etc..

\paragraph{Conclusions.}
The security concern raised in this work is only a \emph{potential} problem for the \emph{long term}. Long term since the quantum computers predicted to be available at least until 2028, would not have enough qubits to run the described Grover's algorithm -- see~\cite{aggarwal17quantum}.%
\footnote{For this reason alone, there is no need to consider some sort of responsible disclosure.} 
The concern is only a potential threat, since we demonstrated that the classical security argument does not hold in the presence of quantum miners. Perhaps security and fairness can be shown by using a different, more elaborate argument or by deploying our proposed countermeasure.

The goal of this work was to establish several facts: (i) There is strong evidence to suggest that, under current Bitcoin rules,  quantum mining would cause a high stale rate. (ii) We propose a simple countermeasure that prevents the high stale rate by prohibiting the AQMS. (iii) Unlike the classical setting, it is not clear what strategy should be suggested as the default (honest) behavior for quantum miners and, more specifically, how many Grover iterations they should apply with our countermeasure. A partial answer to this third question has already been given in~\cite{LRS19}.

We would also like to explain what was \emph{not} shown, and \emph{why}.
This work did not formalize quantum mining as a one-shot game or some other simple model, and analyzed the Nash-equilibria of quantum mining. Additionally, this work did not try to relate between the real world and such a model. To formalize such a model and argue about its properties, it is important to first establish and debate the three facts mentioned in the preceding paragraph to ensure that the model or game that is under analysis captures the desired properties.

%The main open question is to find and understand the properties of the equilibria strategies in the presence of quantum miners, with and without our countermeasure. We hope that in at least one of these cases, the stale rate would be relatively small (for example, less than $5\%$), and that quantum mining would not introduce other severe side-effect to Bitcoin's security and efficiency. 
The main issues that remain to be addressed in terms of Bitcoin and quantum mining are to further understand the equilibria strategies in the presence of quantum miners, beyond the results of~\cite{LRS19}; and to better understand how other aspects of  Bitcoin, such as fairness in the contexts of pool mining~\cite{Ros11,SBBR16}, infiltration attacks~\cite{Eya15}, and selfish mining and its extensions~\cite{eyal14majority,sapirshtein15optimal,nayak16stubborn_mining}, are affected by quantum mining.

\paragraph{Acknowledgments.}
The author is grateful for his fruitful discussions with Yotam Ashkenazi, Ittay Eyal, Robin Kothari, Troy Lee, Maharshi Ray, Yonatan Sompolinsky, Aviv Zohar. He would also like to thank Troy Lee for noticing a vulnerability in the counter-measure presented in an earlier version of this manuscript -- see Appendix~\ref{sec:unsuccessful_countermeasure}.

OS was supported by ERC Grant 280157, by the Israel Science Foundation (grant 682/18), and by the the Cyber Security Research Center at Ben-Gurion University.

\bibliographystyle{alphaabbrurldoieprint}
\bibliography{insecurity_of_bitcoin}
\appendix
\section{Subtle implications of quantum mining}
\label{sec:subtle-impl-quant}
Quantum mining has some subtle implications, which are discussed below. 
\paragraph{Effects on the confirmation time.}
The confirmation time for a transaction is defined as the time it takes for the transaction to be included in a block after being broadcast to the network by the user. Classically, it takes a miner a fraction of a second for each attempt to solve a proof-of-work puzzle. Upon receiving a transaction, a miner can include the transaction in the next attempt to solve the proof-of-work puzzle. Therefore, a block typically contains the transactions paying with the highest fees (measured in bitcoin per byte) that fit into a block\footnote{The size of a block used to be 1 MB and then the calculation was trivial. Recently, Bitcoin upgraded to a Segregated Witness, but the property that a miner selects the transactions that maximize her revenue is still part of the protocol.} at the time that the block was mined. For a user who is willing to pay enough, under normal circumstances (i.e., assuming that the user has internet connection, miners are rational, no denial-of-service attack, etc.), she can guarantee that her transaction will be confirmed in the next block by offering a high enough fee (for example, a higher fee rate than all others). 

A quantum miner can only update her block after a full run of the Grover algorithm. This condition holds with or without our proposed selection rule. For example, if the number of Grover algorithm iterations by all miners is set to $2$ minutes, a user who offers a high enough fee can only guarantee her inclusion in blocks created 2 minutes after the transaction is broadcast. In a more realistic scenario, where the number of Grover iterations is chosen according to some distribution, users who pay high enough fees can only be guaranteed inclusion in the next two blocks (rather than in the next block, which is the current state of affairs).

Apparently, the reduced ability to guarantee inclusion is relatively unimportant since the block creation process is already unpredictable: there are no guarantees regarding the time it will take for the next block to be mined (only the \emph{expected} time is guaranteed). Quantum mining will only increases the unpredictability. More precisely, although classically, a user could guarantee that a transaction would get included in the next block, she could not guarantee the more important property, i.e., \emph{when} it would get included. Because this property cannot be guaranteed, the loss of the guarantees on the less relevant property is of secondary importance. 

\paragraph{Economy of mining equipment.}
Suppose there are two classical mining devices with hash-rates of  $x$ and $2x$.  Other things being equal (such as electricity consumption, etc.), we would expect the second device to cost twice as much as the first, since classically, the revenue from Bitcoin mining is linear in the hash-rate. For quantum devices, the quadratic speed-up renders a different scenario: one would expect, at least naively, that the cost of the second device will be quadruple that of the first.

Were such a cost difference the case, quantum Bitcoin mining hardware manufacturers would be more strongly motivated to improve the hash-rate (analogous to CPU speed), a scenario that may also exist in other markets affected by quantum speed-ups. This example is illuminating, since the connection between computational power and revenue is direct and can be calculated easily. 

\paragraph{Finding the equilibrium and barrier of entry.}
The current strategy for classical miners is extremely simple: mine on top of the tip of the longest chain as fast as possible. It is plausible that in the PQMS equilibrium, the distribution over the number of Grover iterations $Q$ the miner should apply (see Alg.~\ref{alg:quantum_mining_strategies}, line~\ref{line:sample_Q}) in a PQMS would depend on the properties of the other miners (most importantly, the number and hash-rates of the mining devices in each pool). To see a concrete example of such a dependence, see~\cite{LRS19}. This information may not be accessible to all miners, in which case equilibrium would not be achieved.
%\footnote{If the system is in equilibrium, any choice of $Q$ that has support on the distribution is a best response, and therefore, it is easy to achieve.}
Outside equilibrium, miners with more information about the strategies of the other miners would realize higher profits. This may lead to a high barrier to entry, which does not exist now.   

\section{An unsuccessful countermeasure}
\label{sec:unsuccessful_countermeasure}
Eq.~\eqref{eq:new_tie_breaking_rule} provides a new default tie-breaking rule as a countermeasure -- a mechanism that we conjecture prevents the AQMS. In an earlier version of this manuscript, we provided a different, older rule that, it turns out, prevents the AQMS. But in so doing, it introduces a new vulnerability, which was discovered by Troy Lee. The goal of this appendix is to explain the earlier countermeasure and the resulting vulnerability.

The old tie-breaking rule was as follows: Suppose the the tips of the longest chains have timestamps $s_1,\ldots, s_n$, and they were first received at the times $t_1,\ldots,t_n$.  Let $t_{min}=\min_{i\in[n]} t_i$ and let the penalty of each of these tips be defined as $p_i=|t_{min} - s_i|$. The  (old) default strategy was to mine ``on-top'' of the block that had the \emph{lowest} penalty $p_i$.

Though  the (old) tie-breaking rule seems to efficiently prevent the AQMS strategy, consider Mallory, the malicious miner who wants to harm Alice, an honest miner. For Mallory, we present a strategy that has an advantage in terms of knowledge of the network and that can be executed at no cost.

Consider the following example. Let's assume that a block sent by Alice is received by all the other miners after exactly $1$ second and that Mallory is also aware of Alice's activity. Mallory waits for Alice to create a valid block. Suppose that a block  has a timestamp $s_A$ and that it was received by all the other miners at time $t_A = s_A + 1$. Mallory then starts mining a block with the timestamp $s_M=s_A+1$. Suppose that Mallory finds a block after running Grover's algorithm for $100$ seconds. As her block will be received much later than Alice's block -- roughly, $t_A + 100$ -- therefore, $t_{min}$ will remain $t_A$. Note that Mallory's penalty in this case is $|t_{min}-s_M|=|t_{min}-s_A+1|=0$, whereas Alice's penalty will be $|t_{min}-s_A|=1$, and therefore, Mallory will minimize the penalty and win the race. As a result, other miners will start to mine on the top of her block rather than on Alice's block. In terms of costs -- since all miners will mine the top of Mallory's block if she succeeds -- the strategy used by Mallory entails no risk to her, but it may cause Alice to incur significant damage. Insofar as Bitcoin mining is a zero-sum game, it may even be indirectly beneficial to use this strategy not just to attack others. We emphasize that Mallory does not need to be well connected to the other nodes. Rather, she needs to know Alice's propagation time. We are not aware of similar classical or quantum attacks in the context of mining. This attack does not work if the default tie-breaking rule is as defined in Eq.~\eqref{eq:new_tie_breaking_rule}. Note that in this example, $|s_M-t_M|$ is roughly $100$ seconds, and basically, because of that, Mallory's penalty will be higher than that of Alice. Therefore, this specific attack will not work.

\end{document}

%%% Local Variables:
%%% mode: latex
%%% TeX-master: t
%%% End:

%% file: macros.tex
  \theoremstyle{plain}

  \theoremstyle{definition}

  \newtheorem{conjecture}{Conjecture}
  \newtheorem{openproblem}[conjecture]{Open Problem}
  \theoremstyle{remark}
  
  \theoremstyle{plain}
  \newtheorem*{theorem*}{Theorem}
  \newtheorem*{lemma*}{Lemma}
  \newtheorem*{corollary*}{Corollary}
  \newtheorem*{proposition*}{Proposition}
  \newtheorem*{claim*}{Claim}
  
\ifdraft{\newcommand{\authnote}[3]{{\color{#3} {\bf  #1:} #2}}}{\newcommand{\authnote}[3]{}}

\newcommand{\onote}[1]{\authnote{Or}{#1}{green}}

\ifdraft{\linenumbers}{}

\DeclareMathOperator{\E}{\mathbb{E}}